\documentclass[twocolumn,aps,showpacs]{revtex4}
\usepackage {graphicx,graphics}

\newcommand{\3}{$^3$He}

\newcommand{\4}{$^4$He}

\newcommand{\bvn}{\mathbf{v_\mathrm{n}}}

\begin{document}

\title{Quantum and quasi-classical types of superfluid turbulence}

\author{P. M. Walmsley and A. I. Golov}

\affiliation{School of Physics and Astronomy, The University of Manchester, Manchester M13 9PL, UK}

\date{\today}

\begin{abstract}
By injecting negative ions in superfluid \4 in the zero-temperature limit ($T \leq 0.5$~K), we generated tangles of quantized vortex line with negligible large-scale flow. For this {\it quantum} regime of superfluid turbulence, the vortex line length $L$ was found to decay at late time $t$ as $L\propto t^{-1}$; the prefactor being independent of the initial value of $L$. The corresponding effective kinematic viscosity is 0.1~$\kappa$, where $\kappa$ is the circulation quantum. At $T>0.7$~K, a jet of ions generates {\it quasi-classical} tangles identical to those produced by mechanical means. 
\end{abstract}

\pacs{67.25.dk, 67.25.D-, 47.27.Gs}% PACS, the Physics and Astronomy
\maketitle

In turbulent superfluids, flow on short length scales is restricted to quantized vortex lines, each of circulation $\kappa\equiv h/m$ around a narrow core of radius $a_0$, forming a dynamic tangle of the total vortex line length per unit volume $L$ \cite{VinenNiemela2002}. The turbulence can take two very different forms depending on whether the forcing is at scales above or below the mean inter-vortex distance  $\ell\equiv L^{-1/2}$. For flow on scales $> \ell$,  the large quasi-classical eddies are the result of correlations in polarization of vortex lines \cite{Volovik2003}. On the other hand, when forced on quantum scales $< \ell$, the resulting uncorrelated tangle has no classical analogs and should have completely different dynamics first described by Vinen \cite{Vinen1957}. In both cases, the dissipation of flow energy is through the motion of vortex lines; its rate per unit mass being \cite{Vinen2000}
\begin{equation}
	\dot{E} = -\nu(\kappa L)^2.
	\label{EDot}
\end{equation}
 However, the efficiency of the process, expressed through the ``effective kinematic viscosity'' $\nu$, can be different for these two regimes \cite{LNR2007}, which, following \cite{Volovik2003}, we call ``Kolmogorov'' ($\nu_{\rm K}$) and ``Vinen'' ($\nu_{\rm V}$) turbulences. 

In superfluid $^4$He, at high temperatures $T>1$~K, where scattering of thermal excitations (``mutual friction'') dissipates the energy of the tangle and damps the waves on individual vortex lines (Kelvin waves), the values of $\nu_{\rm K}$ and $\nu_{\rm V}$ for both types of turbulence are about the same, $\sim 0.1\kappa$ \cite{Vinen1957,Stalp1999,Vinen2000}. However, at sufficiently low temperatures where the dissipation can only be at very short wavelengths $\ll \ell$ \cite{Vinen2000}, to which the energy can be delivered from larger scales $\sim \ell$ by a cascade of non-linear Kelvin waves \cite{Svistunov1995}, $\nu_{\rm V}(T)$ was so far unknown but $\nu_{\rm K}$ was found to drop to $0.003\kappa$ \cite{WalmsleyPRL2007}. 

To measure the values of $\nu_{\rm V}$ and $\nu_{\rm K}$, one can monitor the free decay of homogeneous tangles. In any tangle, the {\it quantum} energy associated with the quantized flow on length scales $r < \ell$ is $E_{\rm q} = \gamma L/\rho_s$ (per unit mass) \cite{Donnelly1991}, where the energy of vortex line per unit length is $\gamma = B\rho_s\kappa^2$ and $B  \approx  \ln(\ell/a_0)/4\pi$ is approximately constant. If the total energy is mainly determined by $E_{\rm q}$, from Eq.~(\ref{EDot}) we arrive at the late-time decay  \begin{equation}
	L = B \nu_{\rm V}^{-1}t^{-1}.
	\label{L-t-V}
\end{equation}
The typical values of $\ell$ are 0.1--1~mm. For \4, $\kappa_4 = h/m_4 = 1.00\times10^{-3}$~cm$^2$/s and $a_0 \sim 0.1$~nm \cite{Donnelly1991} making $B \approx 1.2$, while for \3-B, $\kappa_3 = h/2m_3 = 6.6\times10^{-4}$~cm$^2$/s, $a_0 \sim$ 13--65~nm \cite{helium3} and $B \approx 0.7$.

Now suppose that the tangle is structured due to the presence of flow on classical scales $r>\ell$, and the additional energy of this {\it classical} flow $E_{\rm c}$ is much greater than $E_{\rm q}$. For the Kolmogorov spectrum between wavenumbers $k_1$ and $k_2$ ($k_1 \ll k_2$), while the size of the energy-containing eddy stays equal to the size of container $d$ (i.~e. $k_1 \approx 2\pi/d$), the late-time decay becomes \cite{Stalp1999,Stalp2002}
\begin{equation}
	L = (3C)^{3/2}\kappa^{-1}k_1^{-1}\nu_{\rm K}^{-1/2} t^{-3/2}. 
	\label{L-t-K}
\end{equation}
The experimental value for the Kolmogorov constant for classical homogeneous isotropic turbulence is $C \approx 0.5\times \frac{55}{18} = 1.5 (\pm 11\%)$ \cite{Sreeni1995}. All absolute values of $\nu_{\rm K}$ quoted in this paper are obtained from the fit to Eq.~(\ref{L-t-K}) using $C=1.5$ and $k_1=2\pi/d$.    

In intermediate cases, the initial transient (before $L(t)$ takes the form of either Eq.~(\ref{L-t-V}) or Eq.~(\ref{L-t-K}), depending on initial conditions) can look much like $L \propto t^{\epsilon}$ with $\epsilon \approx -1$ \cite{Stalp1999}. However, the absolute value of the prefactor in the fit $L \propto t^{-1}$ will depend on the initial level of pumping and will never drop below $B\nu^{-1}_{\rm V}$. 

\begin{figure}[h]
\centerline{\includegraphics[width=8cm]{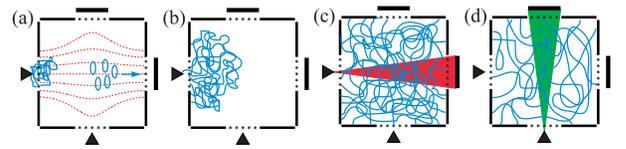}}
\caption{(color online) Cartoon of the vortex configurations (side view) at different stages.  The shaded areas indicate the trajectories of ions used to probe the tangle along two orthogonal directions. To avoid contamination of the developing tangle with new ions and vortices, only one probe pulse of ions was fired for each realization of the tangle. (a) At $t=0$~s, a pulse of CVRs is injected from the left injector. While most make it to the collector as a sharp pulse, some got entangled near the injector. (b) At $t\sim 5$~s, the tangle spreads into the middle of the cell. (c) At $t\sim 20$~s, the tangle has occupied all volume. (d) For up to 1000~s, the homogeneous tangle is decaying further.} \label{fig1}
\end{figure}

\begin{figure}[h]
\centerline{\includegraphics[width=7cm]{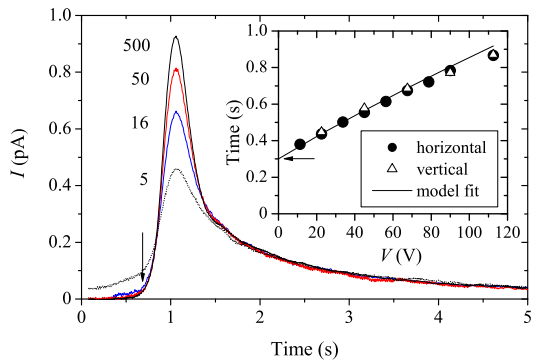}}
\caption{(color online) Transients of the current of CVRs to the collector after different waiting times between the pulses (shown in seconds); the duration of injection is 0.3~s, $V = 90$~V, $T=0.15$~K. The arrival time (leading edge indicated by an arrow), 0.7~s, is the same for all pulses. The inset shows the dependence of the arrival time on the driving voltage.} \label{fig3}
\end{figure}

We generated turbulence in a cube-shaped container with sides $d=4.5$~cm filled with \4 at $p=0.1$~bar by injecting  negative ions (electrons in a bubble state) through the center of a square plate (Fig.~1). There were two field-emission injector tips (``left'' and ``bottom'') and two collectors on opposite sides (``right'' and ``top'') \cite{IonCellPobell}. The driving field of mean value $V/d$, shown in Fig.~1(a), was maintained by applying a potential $-V$ to the injector plate and $-V/2$ to all side plates relative to the collector grid. The turbulence was detected by scattering short time-resolved pulses of free ions at $T>0.7$~K or charged vortex rings (CVR) at $T<0.7$~K off the vortex tangle \cite{WalmsleyPRL2007}. The ion-vortex trapping diameter $\sigma$ for horizontal injection-collection was calibrated {\it in situ} on vertical vortex arrays of known density at steady rotation of the cryostat \cite{WalmsleyPRL2007}.

At $T<0.7$~K, each injected electron dresses itself in a quantized vortex ring, somewhere between the tip and the nearby grid, and then propagates along with it into the cell. The ring energy $E_r = \frac{1}{2}\kappa^2\rho_s R (\Lambda_{\rm r}-2)$ and velocity $v_r \approx \frac{\kappa}{4\pi R}(\Lambda_{\rm r}-1)$ depend solely on its radius $R$ (here $\Lambda_{\rm r} = \ln(8R/a_0) \approx 11$) \cite{Donnelly1991}. The arrival of pulses of current carried by ballistic CVRs is shown in Fig.~2. The extrapolation of the dependence of the time-of-flight on the driving voltage to $V=0$ gives 0.3~s (inset) corresponding to CVRs injected through the grid with initial velocities $v_0\approx 15$~cm/s, radius $R_0\approx 0.53$~$\mu$m and energy 21~eV. Even though the side and bottom tips had very different threshold voltages for electron injection, 300 and 120~V, these initial radii of the rings were about the same for both.
No dependence of the ring energy on the injected current in the range $10^{-12}$ -- $10^{-10}$~A was found. The mutual friction $\alpha$ limits the range of a ballistic ring to $R_0/\alpha$ \cite{Donnelly1991}. For CVRs to survive the distance to collector $d$ one needs $\alpha < R_0/d \sim 10^{-5}$ (corresponding to $T<0.5$~K), although a propelling force due to the driving field extends this temperature range. 

\begin{figure}[h]
\centerline{\includegraphics[width=7cm]{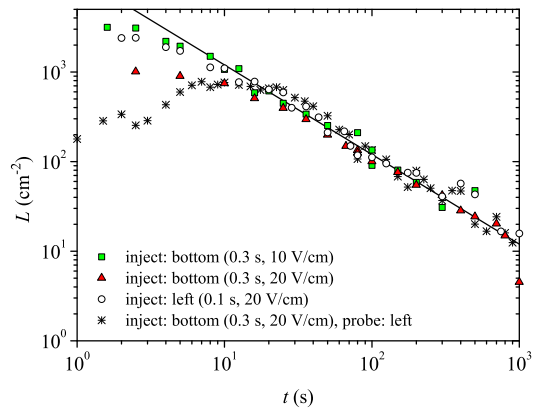}}
\caption{(color online) Free decay of a tangle at $T=0.15$~K. The injection direction and duration, and driving field are indicated. Probing with pulses of CVRs of duration 0.1--0.3~s were done in the same field as the initial injection, and also in the same direction except in one case ($*$).  The line $L\propto t^{-1}$ corresponds to Eq.~(\ref{L-t-V}) with $\nu_{\rm V} = 0.1\kappa$.}
\label{fig3}
\end{figure}

When such rings collide, they build a tangle which then spreads into all volume. The decrease in the peak height $I$ of the current pulses due to the ballistic CVRs (shown in Fig.~2), relative to the long time amplitude $I(\infty)$ when all vortices have decayed, is the measure of blocking the paths of some ballistic CVRs with the probability per unit length $L \sigma$ where $\sigma \sim R$. We hence determine the vortex density after time $t$ since stopping injection as $L(t) = (\sigma d)^{-1}\ln\frac{I(\infty)}{I(t)}$. The slowly decaying component of the current (i.e. of the decay time much longer than the pulse width 0.1--0.3~s and the time constant of the amplifier 0.15~s), that appears in Fig.~2 as ``tails'' after each pulse at times $t>1.5$~s, is due to the ions trapped on entangled vortices originating after each pulse. Their contribution, clearly seen in Fig.~2 as being of $\sim 0.04$~pA after only 5~s of waiting between pulses, was subtracted when determining $I(t)$. The free decay for these tangles $L(t)$ at $T<0.5$~K is shown in Fig.~3, which is our central result. We found that the late-time decay is insensitive to the details of generating the initial tangle such as injection duration (0.1 -- 1~s) and intensity ($10^{-12}$ -- $10^{-10}$~A), and driving field (0 -- 20~V/cm) (see Fig.~3), i.e. the decay curve is {\it universal} and follows Eq.~(\ref{L-t-V}). This gives strong support to the interpretation that the dynamics is that of the random tangle and not just the transient of a structured tangle with a not yet saturated energy-containing length.
No changes with temperature were observed at $T\leq 0.5$~K. The values of $\nu_{\rm V}/\kappa$, obtained using Eq.~(\ref{L-t-V}), were slightly different for $L(t)$ probed in horizontal ($0.083\pm 0.004$) and vertical ($0.120\pm 0.013$) directions. This might reflect the fact that trapping diameters $\sigma$ were slightly different for CVRs injected by different tips. As we could only calibrate $\sigma$ for the horizontal direction, and then used the obtained value to quantify $L$ for both directions, the absolute value of $\nu_{\rm V} = 0.08 \kappa$ seems more reliable. Note that, because of the finite width of the probe pulses 0.1--1.0~s and recorded time of flight $\sim$0.15 -- 1.0~s for CVRs and free ions, the decay time $t$ has an uncertainty of 0.1--1~s; this is negligible for points at $t>20$~s that were used for quantitative analysis. 

During steady injection, the slow component of the collector current, associated with the spread of the tangle containing the trapped ions, only arrived after $\sim 10$~s. This implies that the initial tangle is always created near the injector and not near the collector at the opposite side of the cell. To study the dynamics of how the tangle spreads out into space, we probed $L(t)$ {\it across} the direction of injection (Fig.~3($*$)): within some 5--10~seconds the tangle reaches the center of the cell, then fills all volume and becomes homogeneous (as $L(t)$ becomes indistinguishable from those measured along the direction of injection) after $\sim 20~s$. The dynamics of spreading was found to be independent of the driving field. The observed time, $\sim 20~s$, is surprisingly short if one compares it with the estimate $d^2/D \sim 2\times 10^5$~s based on the simulated value of the diffusion coefficient \cite{TsubotaPhysica2003} $D\sim 0.1\kappa$ for $L$. However, if we consider ``evaporation of vortex loops'' from the boundary regions of strongly inhomogeneous tangles, the rate of spreading is comparable with $\sim 1$~mm/s observed in simulations \cite{BarenghiEvaporation}. Also, the initial tangle might maintain a certain polarization as a memory of the orientation of the initial injected CVRs that it was created from. This would speed up the process of spreading before the tangle occupies all volume.

The field due to the trapped charge was found not to affect the dynamics of the tangle's decay. This was checked for various ratios of the total injected charge of density $n$ to $L$ up to $n/L \sim 10^5$~cm$^{-1}$, as well as in different driving fields between $-5$~V/cm and $20$~V/cm. As the time of flight of CVRs (Fig.~2) was independent of the trapped charge (but dependent on the driving voltage) the field due to the trapped charge was always much smaller than the driving one.

\begin{figure}[h]
\centerline{\includegraphics[width=7cm]{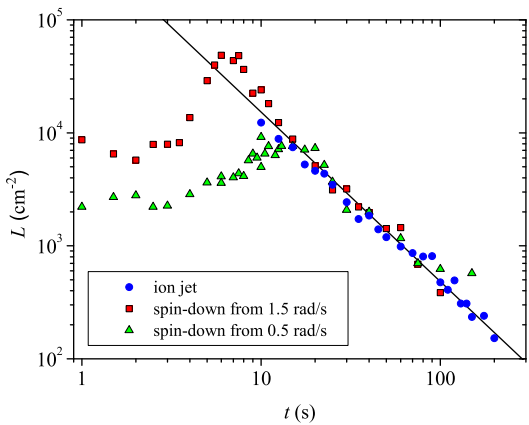}}
\caption{(color online) Free decay of a tangle produced by a jet of ions ($\bullet$) from the bottom injector into a 10~V/cm field for 150~s, as well as by an impulsive spin-down to rest \cite{WalmsleyPRL2007} from 1.5~rad/s and 0.5~rad/s, at $T=1.60$~K. All tangles were probed by pulses of free ions in the horizontal direction (spin-down data: probe field 20~V/cm, pulse length 0.5~s; ion jet data: probe field 10~V/cm, pulse length ~1.0~s ). The line $L\propto t^{-3/2}$ corresponds to Eq.~(\ref{L-t-K}) with $\nu_{\rm K} = 0.2\kappa$.}
\label{fig4}
\end{figure}

Yet, after a sufficient injection time the classical energy $E_{\rm c}$ can become dominant. Indeed, the observed late-time decay at $T>0.7$~K (Fig.~4), $L \propto t^{-3/2}$, especially prominent for dense initial tangles after long ($>10$~s) injection, was virtually identical to that of turbulence produced by mechanical means.
Using Eq.~(\ref{L-t-K}), we extract the values of $\nu_{\rm K}(T)$ for $L\propto t^{-3/2}$ and plot them in Fig.~5 for ion-jet, spin-down and grid turbulences \cite{Stalp2002}. The spin-down data include those published \cite{WalmsleyPRL2007} and new points. The revised fitting routine, in which we only fit the $L(t)$ for $t$ greater than at least 50~s (to ensure that the transient behavior is not mistaken for the late-time decay), has resulted in slightly smaller the values of $\nu_{\rm K}$ at $T$ near 0.8--1.0~K than those published in \cite{WalmsleyPRL2007}. 

\begin{figure}[h]
\centerline{\includegraphics[width=7.5cm]{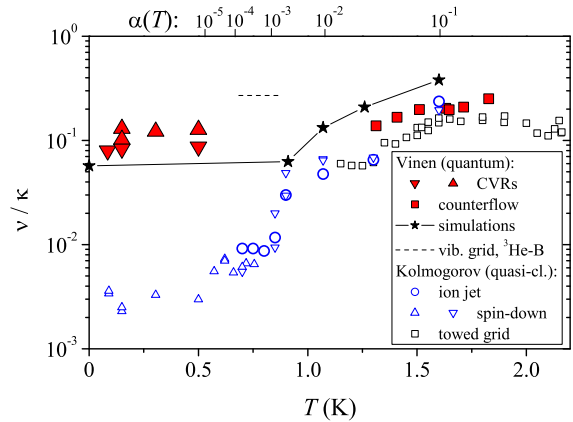}}
\caption{(color online) The effective kinematic viscosity for different types of superfluid turbulence.  Vinen (quantum) type, generated by: CVRs (this work,  $L(t)$ measured along the vertical ($\bigtriangleup$) and horizontal ($\bigtriangledown$) directions), build-up of counterflow \cite{Vinen1957}, simulations \cite{Tsubota2000}, vibrating grid in $^3$He-B vs. temperatures at which \4 has comparable values of $10^{-4} < \alpha < 2\times 10^{-2}$ \cite{Bradley2006}. Kolmogorov (quasi-classical) type: ion jet (this work), spin-down \cite{WalmsleyPRL2007} sampled by CVRs ($\bigtriangleup$) and free ions ($\bigtriangledown$), towed grid \cite{Stalp2002}.}  
\end{figure}

Let us now discuss the temperature dependence of $\nu_{\rm V}(T)$ (Fig.~5). 
For $T=0$, a tangle of vortex filaments initiated at short lengths and then allowed to decay through frequent reconnections of vortices was simulated using the local induction approximation \cite{Tsubota2000}, and the $L\propto t^{-1}$ decay was observed yielding $\nu_{\rm V} = 0.06 \kappa$. Another theoretical simulation, of the decay of a vortex ring through sound emission by Kelvin waves without any reconnections gives only $\nu_{\rm V} = 1.1 \times 10^{-3} \kappa$ \cite{BarenghiDecay} thus emphasizing the importance of reconnections. Experimentally, only inhomogeneous tangles were generated in the $T=0$ limit by colliding vortex rings before this work: by a jet of \4 through an orifice \cite{Guenin}, injection of electrons \cite{McClintock1} and by vibrating objects such as a grid in \3-B \cite{Bradley2006}. Only the latter was studied quantitatively and revealed that within a range of levels of initial pumping, the free-decay transients (before switching to late-time decay $L\propto t^{-3/2}$) of type $L\propto t^{-1}$ were observed with a prefactor increasing with increasing initial drive. Applying Eq.~(\ref{L-t-V}) to the lowest one hence yields the lower margin of $\nu_{\rm V} = 0.27\kappa_3$ for \3-B which is factor 2--3 larger than our $\nu_{\rm V}$ for \4. Of corse, the dissipation in $^3$He-B might have additional channels \cite{VinenNiemela2002}.

At $T>1$~K ($\alpha > 10^{-2}$), computer simulations of tangle's decay while enforcing $\bvn=0$ \cite{Tsubota2000} yield an increase in $\nu_{\rm V}(T)$ with temperature. The experimental $\nu_{\rm V}$, obtained for forced counterflow in narrow channels \cite{Vinen1957}, are systematically lower than the simulated ones. With increasing temperature the normal component becomes less viscous, hence it becomes impossible to enforce the $\bvn=const$ condition. Yet, approaching $T\sim 1$~K from above (as the normal component firstly becomes laminar and eventually irrelevant) the high-temperature data for $\nu_{\rm V}$ seem to converge to the $T=0$ value $\sim 0.1\kappa$. 

There exist two models for the drop in $\nu_{\rm K}$ upon approaching the $T=0$ limit. L'vov {\it et al.} \cite{LNR2007} argue that conversion of the quasi-classical energy into Kelvin waves has constraints which cause a pile-up of vorticity at scales $>\ell$ and hence a decrease in $\nu_{\rm K}$. As Vinen tangles do not possess any energy at large scales, this scenario is consistent with $\nu_{\rm V}(T)$ having no similar drop. Alternatively, Kozik and Svistunov \cite{KS2008} explained $\nu_{\rm K}(T)$ for $T<1$~K as the result of fractalization of the tangle at smaller length scales $<\ell$ as $\alpha(T)$ decreases between $\sim 10^{-2}$ and $10^{-5}$. To interpret the observed weak temperature dependence of $\nu_{\rm V}(T)$ within this approach one has to assume that, for Vinen tangles, fractalization does not add a significant extra contribution to $L$ for $T < 1$~K. 

To conclude, we generated a homogeneous quantum (Vinen) vortex tangle in the $T=0$ limit and found that, independently of the level of initial forcing, its late-time decay follows the law $L\propto t^{-1}$ with a universal prefactor within more than two orders of magnitude. We hence extracted the effective kinematic viscosity for such a tangle, $0.08 \kappa$, in reasonable agreement with the value $0.06\kappa$ from simulations of the reconnections-driven decay \cite{Tsubota2000}, but about 30 times larger than that for a homogeneous quasi-classical (Kolmogorov) turbulence (generated by either impulsive spin-down to rest \cite{WalmsleyPRL2007} or by ion jet in this work). 
%Amazingly, at $T=0$  quantum tangles decay basically as fast as at $T=1$~K. 
Thus, in the zero temperature limit the energy flux fed at quantum length scales $< \ell$ generates little excess vortex length unlike the one fed at classical scales $>\ell$. 

We acknowledge fruitful discussions with Henry Hall and Joe Vinen, and the contributions of Alexandr Levchenko and Steve May in the construction and improvement of the apparatus. Support was provided by EPSRC under GR/R94855 and EP/E001009.


\begin{thebibliography}{99}
\bibitem{VinenNiemela2002} W. F. Vinen, J. J. Niemela, J. Low Temp. Phys. {\bf 128}, 167 (2002).
\bibitem{Volovik2003} G. E. Volovik, JETP Lett. {\bf 78}, 533 (2003). 
\bibitem{Vinen1957} W. F. Vinen, Proc. Roy. Soc. London, Ser. A {\bf 240}, 114, 128 (1957); {\bf 242}, 493 (1957); {\bf 243}, 400 (1958). 
\bibitem{Vinen2000} W. F. Vinen, Phys. Rev. B {\bf 61}, 1410 (2000). 
\bibitem{LNR2007} V. S. L'vov {\it et al.}, Phys. Rev. B {\bf 76}, 024520 (2007).
\bibitem{Stalp1999} S. R. Stalp {\it et al.}, Phys. Rev. Lett. {\bf 82}, 4831 (1999).
\bibitem{Svistunov1995} B. V. Svistunov, Phys. Rev. B {\bf 52}, 3647 (1995).
\bibitem{WalmsleyPRL2007} P. M. Walmsley {\it et al.}, Phys. Rev. Lett. {\bf 99}, 265302 (2007). 
\bibitem{Donnelly1991} R. J. Donnelly, {\it Quantized Vortices in Helium
II}, Cambridge University Press 1991.
\bibitem{helium3} V. B. Eltsov {\it et al.}, Prog. Low Temp. Phys., Vol. XV, ed. W. P. Halperin (Elsevier
B.V., Amsterdam), 1 (2005). 
\bibitem{Stalp2002} S. R. Stalp {\it et al.}, Phys. Fluids {\bf 14}, 1377 (2002).
\bibitem{Sreeni1995} K. R. Sreenivasan, Phys. Fluids {\bf 7}, 2778 (1995). 
\bibitem{IonCellPobell} P. M. Walmsley {\it et al.}, J. Low Temp. Phys. {\bf 146}, 511 (2007). 
\bibitem{TsubotaPhysica2003} M. Tsubota {\it et al.}, Physica B {\bf 329}, 224 (2003).
\bibitem{BarenghiEvaporation} C. F. Barenghi {\it et al.}, Phys. Rev. Lett. {\bf 89}, 155302 (2002). 
\bibitem{Tsubota2000} M. Tsubota {\it et al.}, Phys. Rev. B {\bf 62}, 11751 (2000).
\bibitem{BarenghiDecay} M. Leadbeater {\it et al.}, Phys. Rev. A {\bf 67}, 015601 (2003). 
\bibitem{Guenin} B. M. Guenin and G. B.  Hess, J. Low Temp. Phys. {\bf 33}, 243 (1978). 
\bibitem{McClintock1} R. M. Bowley {\it et al.}, Phil. Trans. R. Soc. Lond. A {\bf 307}, 201 (1982). 
\bibitem{Bradley2006} D. I. Bradley {\it et al.}, Phys. Rev. Lett. {\bf 96}, 035301 (2006). 
\bibitem{KS2008} E. V. Kozik and B. V. Svistunov, Phys. Rev. B {\bf 77}, 060502 (2008); Phys. Rev. Lett. {\bf 100}, 195302 (2008).
\end{thebibliography}
\end{document}